\documentclass[11pt,toc]{JHEP}
\usepackage{cite}
\catcode `@=11 \@addtoreset{equation}{section} \catcode `@=12
\def\be{\begin{equation}}
\def\ee{\end{equation}}
\def\ba{\begin{array}{l}}
\def\ea{\end{array}}
\def\bea{\begin{eqnarray}}
\def\eea{\end{eqnarray}}

\def\nn{\nonumber\\}

\def\eq#1{(\ref{#1})}

\def\gap#1{\vspace{#1ex}}
\def\del{\partial}

\title{\Large Aspects of Semiclassical Strings in AdS$_5$}
\author{
Gautam Mandal${}^\dagger$, Nemani V. Suryanarayana${}^\ddagger$ and 
Spenta R. Wadia${}^\dagger$\\ 

\gap2

$\!\!{}^\dagger$Tata Institute of Fundamental Research \\
\, Homi Bhabha Road, Mumbai 400 005, India\\
~~\\
$\!\!{}^\ddagger$DAMTP, Centre for Mathematical Sciences, \\
\, Wilberforce Road, Cambridge CB3 0WA, UK \\

\gap1

e-mail: \email{mandal@theory.tifr.res.in, v.s.nemani@damtp.cam.ac.uk,
wadia@theory.tifr.res.in}
}

\preprint{
\hepth{0206103}
\\ TIFR/TH/02-20 \\ DAMTP-2002-71
}

\abstract{We find an infinite number of conserved currents
and charges in the semiclassical limit $\lambda
\to \infty$ of string theory in AdS$_5 \times S^5$
and  remark on their relevance to conserved charges in the dual gauge
theory. We establish a general procedure of exploring the
semiclassical limit by viewing the classical motion as collective
motion in the relevant part of the configuration space. We illustrate
the procedure for semiclassical expansion around solutions of string
theory on AdS$_5 \times (S^5/Z_M)$.}

\keywords{WKB, String, AdS-CFT}

\begin{document}

\section{Introduction}
Recently string propagation in pp-wave backgrounds has attracted much
interest \cite{blaua, blaub, blauc}. One of the attractive features of
this background is that the tree level string theory is exactly
soluble in the Green-Schwarz formalism \cite{metsaev, metsyet}. In a
parallel development Berenstein, Maldacena and Nastase (BMN)
\cite{bmn} presented a derivation of the tree level string theory
starting from the dual SYM theory. In fact despite various efforts
over the past 25 years \cite{thooft, polyakov, makeenko, spenta,
largeN, mal97,Gubser:1998bc, Witten:1998qj, Maldacena:1998im,
Witten:1998zw, Polyakov:2000ti} this is the first time a precise
derivation of critical string theory (in a given background) has been
given from a gauge theory. We also see how a spacetime background
arises from a gauge theory.

Subsequently there has been work along the lines of BMN in
generalizing to strings propagating in various spaces related to
$AdS_5\times S^5$ \cite{bgmnn} where a similar analysis was done for
open strings. For other related developments see \cite{ikm, gomoogu,
russotsey, pandosonn, alishajabba, sjrey, takayanagi, kehagi, ohtatar,
russo, mukund,spradvolov, staudacher, bn, gross, shiraz, rajesh,
chuholi, kikilee}.

In another interesting development Gubser, Klebanov and Polyakov (GKP)
\cite{gkp} proposed to explore time dependent soliton solutions of the
sigma model on $AdS_5\times S^5$ and their relation to specific
sectors of the dual gauge theory. For example a collapsed string
moving on the great circle of $S^5$ with angular velocity $\omega$
corresponds as expected to a chiral primary state in the gauge theory
with a large value of the R-charge: $J=\sqrt{\lambda}\,\omega$, where
$\lambda$ is the 'tHooft coupling: $\lambda=g_{YM}^2N$. In the
semi-classical limit we send $N\rightarrow\infty$, holding $g_{YM}^2$
small and fixed. They also explored various other soliton solutions
corresponding to a spinning string in $AdS_5$. These solutions were
generalised in \cite{frotsey,russo,cernbh}. These solitons do not
correspond to chiral primary states and the precise theory needs
further exploration.

Our work is an attempt to explore further the GKP proposal. There are
two aspects to this paper. First, a formal aspect points out that the
sigma models under question have an infinite number of classical
conserved currents \cite{Pohlmeyer:1975nb, Luscher:1977rq, Brezin:am}.
We point out the relevance of the conserved charges to the classical
string theory. 

The second aspect of the paper deals with the orbifold $AdS_5\times
S^5/Z_M$. The Penrose limit of this model has been discussed in
\cite{alishajabba} and the gauge theory correspondence has been
studied in \cite{mrv}. We detail the analysis of the fluctuations
around the soliton in $AdS_5\times S^5/Z_M$ and reproduce the string
propagating in a pp-wave geometry with a compact light like circle. In
doing this analysis we develop a general procedure of exploring the
semiclassical limit by viewing the classical motion as collective
motion in the relevant part of the configuration space.

Below, in Section 2, we review the relevant aspects of the string
sigma model on $AdS_5\times S^5$ background
\cite{Metsaev:1998it,Berkovits:2000fe}. In Section 3 we turn to the
conserved currents of the non-linear sigma model following
\cite{Luscher:1977rq} and discuss their relevance to string theory for
worldsheet solitons. In Section 4, we present a solitonic solution of
the sigma model on the orbifold $AdS_5\times S^5/{Z_M}$ and describe
its quantization in a semiclassical expansion using the method of
collective coordinates. The spectrum corresponds to that of a winding
string propagating in the pp-wave geometry of
\cite{alishajabba,mrv}.

\section{String theory on AdS$_5 \times S^5$}

Here we start with a brief review of the relevant aspects of string
sigma model on $AdS_5 \times S^5$ background. We represent the metric
on $AdS_5 \times S^5$ by the line element
\be 
ds^2 = R^2\left(
dN_\alpha dN^\alpha + dn_a dn^a \right)
\label{metric-0}
\ee
where $N^\alpha$ represents a ``timelike'' unit
vector in $R^{2,4}$:
\be
- \eta_{\alpha\beta} N^\alpha N^\beta = N_0^2 + N_5^2 - 
N_1^2 - N_2^2 - N_3^2 - N_4^2 = 1
\label{unit-N}
\ee
and $n_a$ is a unit vector in $R^6$:
\be
n_1^2 + n_2^2 + n_3^2 + n_4^2 + n_5^2 + n_6^2 = 1.
\label{unit-n}
\ee

The bosonic part of the world-sheet action of string
theory in the above space is \cite{Metsaev:1998it,Berkovits:2000fe}
\begin{equation}
S = -\frac{R^2}{4\pi\alpha'} \int d^2\xi\ \sqrt{g} g^{\mu\nu}
\left[
\del_\mu N^\alpha \del_\nu N_\alpha +
\del_\mu n^a \del_\nu n^a
\right]
\label{action0}
\ee
where we have omitted the fermionic terms (which include the RR
couplings). The fermionic terms are of course crucial for a consistent
quantum string theory. However, it is clear that for classical
solutions of the string world-sheet action, it suffices to set the
fermions to zero and consider only the bosonic terms in the action. We
have also chosen to work with a Minkowski world sheet. As in \cite{gkp}
we will use the notation
\[
\alpha = \alpha'/R^2
\]
and understand $\alpha\to 0$ as the classical limit. The 'tHooft
coupling of the dual gauge theory is given by 
\[\lambda \sim R^4/(\alpha')^2 = 1/\alpha^2.
\]

After fixing the conformal gauge on the metric \cite{Metsaev:1998it} 
$\sqrt{g} g^{\mu\nu}= \eta^{\mu\nu}= {\rm Diag}[-1,1]$, the
action becomes a sum of non-linear sigma models on $S^5$ and
AdS$_5$ in two-dimensional Minkowski space:
\be
S = -\frac{R^2}{4\pi\alpha'} \int d^2\xi\ \left[
\del_\mu N^\alpha \del^\mu N_\alpha +
\del_\mu n^a \del^\mu n^a
\right]
\label{action}
\ee
which comes together with two constraints $T_{++}= T_{--}=0$.
These read
\bea
\del_+N^\alpha\del_+ N_\alpha +
\del_+n^a\del_+ n^a 
&&=0
\nn
\del_-n^a\del_- n^a + \del_- N^\alpha\del_-N_\alpha &&=0
\label{constraint}
\eea
In the above equations, $\del_\pm = \frac{\del}{\del \sigma^\pm},
\, \sigma^\pm = \frac12 (\tau \pm \sigma)$.
The action \eq{action} leads to the following equations
of motion:
\bea
\del^2 n^a &&= n^a n^b \del^2 n^b,
\label{n-eom}\\
\del^2 N^\alpha &&= -N^\alpha N_\beta \del^2 N^\beta
\label{N-eom}
\eea 
We should note that at the classical level, the system
\eq{action} is conformally invariant.

\section{Infinite number of conserved currents}

Luscher and Pohlmeyer \cite{Luscher:1977rq}
(see also \cite{Brezin:am}) show that for a classical
non-linear sigma model, namely one that satisfies
\eq{unit-n} and \eq{n-eom},  there are infinite 
number of (non-local) conserved currents $J^n_\mu$ in the system:
\be
\del_\mu J_n^\mu=0
\label{del-j}
\ee
The discussion in the above papers assumes that the
NLSM is defined on a plane while the string theory world-sheet
is usually a cylinder. This is not a problem; since
the theory classically is conformally invariant, we can 
make a conformal transformation from a cylinder to
a plane to use the results of \cite{Luscher:1977rq,Brezin:am}.

These currents are constructed recursively. From
a $J_n^\mu$ satisfying \eq{del-j}, it is always
possible to define $\chi_n$ so that 
\be
J_{n,\mu}=: \epsilon_{\mu\nu} \del^\nu \chi_n
\ee
Define, further,
\be
(A_\mu)_{ab}= n_a\del_\mu n_b - n_b \del_\mu n_a
\ee
Then, the $n+1$-th current is
\be
J_{n+1,\mu}:= (\del_\mu + A_\mu) \chi_n
\ee
Note that equation \eq{n-eom} implies
$\del^\mu A_\mu=0$. Thus we can start with $J^1_\mu
= A\mu, (\chi_0= 1)$ and so on.

A similar remark goes through also for the non-linear sigma-model
based on AdS$_5$ and one can again, independently, construct an
infinite number of conserved currents.

At this point, one must note that the equations of motion \eq{n-eom}
and \eq{N-eom} are constrained together by \eq{constraint}. Then how
does our above construction of infinite number of conserved currents
remain valid? After all, we are doing string theory here as opposed to
a non-linear sigma model.

This turns out to have an easy answer.  The constraints are first
class, so they can in principle be gauge-fixed. However, even without
gauge fixing, it is easy to see that the equations of motion will be
always valid (it only amounts to redundancy in the description of
dynamical evolution). Thus, the construction of the conserved currents
that we have described above remains valid.  Indeed, the argument can
be made differently. Suppose we have fixed the gauge; then on the
gauge-fixed surface also the equations of motion will remain valid,
and hence the construction of the conserved currents. Thus
we do have an infinite number of conserved currents in string 
theory.

\subsection{Spacetime significance of these charges}

We found above that there are infinite number of classically conserved
currents in the AdS$_5 \times S^5$ string theory.  How about {\em
charges}?  For Minkowski signature of the world-sheet, it is easy to
fix a partial gauge $\tau=t$. In this case the infinite number of
conserved currents gives rise to an infinite number of conserved
charges, which are conserved also in time in the {\em target space}.
We note that typically global charges in the world-sheet theory
correspond to a local symmetry in the target space, which, in
turn, corresponds to global charges in the dual gauge theory.

The above discussion would imply that the dual ${\cal N}=4$ super Yang
Mills theory has an infinite number of conserved, perhaps nonlocal,
charges in the classical limit $\lambda\to \infty$ (note that the time
in the string theory is the same as that of the gauge theory). In this
paper we will not discuss the rather important question of what
happens to these conserved charges at $\lambda < \infty$, in
particular whether the conservation laws have anomalies. This issue is
somewhat more subtle in string theory compared to usual NLSM because
of the inclusion of fermionic terms in the currents in the quantum
theory. It is also important to note that the existence of such
integrable structures has been hinted at in \cite{wit-talk} at the
opposite limit $\lambda \to 0$.  It would be obviously interesting to
unravel any such integrable structure at a general $\lambda$ (see
Figure 1).

\hspace{10ex}
\vbox{
\begin{picture}(200,100)(1,1) 
\put(1,41){$\lambda=0$}
\put(201,41){$\lambda=\infty$}
\put(101,41){$\lambda$}
\put(11,31){$\bullet$}
\put(11,33){ \vector(1,0){190}}
\put(1,19){Integrable}
\put(201,19){Integrable}
\put(101,19){\bf ?}
\end{picture}

\gap1

\ ~~~~~~~~~~~~~~~~~~~~~ Figure 1 \hfill\break
}

\gap5

\subsection{AdS$_3 \times S^3$}

{}From the viewpoint of integrability, string theory on AdS$_3 \times
S^3$ appears to be easier. Since $S^3$ is a group manifold, more is
known about its integrable structure and exact solutions. Besides, it
is also known that there are additional world-sheet charges which are
local \cite{evans}. Furthermore, there seems to be a rather direct
correspondence between the current algebra on the world-sheet and a
corresponding current algebra in spacetime \cite{giveon}. We hope to
return to these issues and its significance for the dual CFT in a
future publication.

\section{String Solitons on $AdS_5 \times S^5/{Z_M}$}
\label{orbifold}

In the second half of the paper we want to report on semiclassical
solutions of string theory on $AdS_5\times S^5/{Z_M}$ orbifold.  Part
of our motivation is to share some insights on the general method of
dealing with such semiclassical limits which applies to more general
manifolds and of course to the original AdS$_5\times S^5$ string
theory as well.  We seek a solution of the sigma model, in the spirit
of \cite{gkp}, which sees the pp-wave geometry of
\cite{alishajabba}. The dual gauge theory was discussed in
\cite{mrv}. 

We recall that the metric of $AdS_5 \times S^5$ is given
by \eq{metric-0},\eq{unit-n} and \eq{unit-N}. We will use below the
following explicit representation for the unit vector $N^\alpha$:
\begin{eqnarray}
\label{nis}
N^0 &=& \hbox{cosh}\rho\, \hbox{cos}t \cr
N^5 &=& \hbox{cosh}\rho\, \hbox{sin}t \cr
N^j &=& \hbox{sinh}\rho\, \Omega^j ~~~~ j = 1,..,4
~~~~ \hbox{such that} ~~~~
\Omega^j\Omega^j =1 
\end{eqnarray}
With this,
\be
ds^2_{AdS_5} = R^2 [ -\hbox{cosh}^2\rho\, dt^2 + d\rho^2 +
\hbox{sinh}^2\rho\, d\Omega^2] 
\ee
We will further use the following representation for 
the unit vector $n_a$
\begin{eqnarray}
\label{zeds}
n_1 + i\,n_2 &&= z_1 = \,\hbox{sin}\beta\, e^{i\theta} \cr
n_3 + i\,n_4 &&= z_2 = \,\hbox{cos}\beta\,\hbox{cos}\gamma\, e^{i\chi} \cr
n_5 + i\,n_6 &&= z_3 = \,\hbox{cos}\beta\,\hbox{sin}\gamma\,e^{i\phi}
\end{eqnarray}
With this, the metric of  $S^5$ becomes:
\be
ds^2_{S^5} = R^2[d\beta^2 + \hbox{sin}^2\beta\,
d\theta^2 + \hbox{cos}^2\beta\, (d\gamma^2 + 
\hbox{cos}^2\gamma\, d\chi^2 + \hbox{sin}^2\gamma\,d\phi^2 ) ]
\ee

Now we come to AdS$_5 \times (S^5/Z_M)$.
The $Z_M$ orbifold action on $S_5$ is defined by
\be
z_1 \to z_1, z_2 \to e^{2\pi i/M} z_2, z_3 \to e^{-2\pi i/M} z_3
\label{zequiv}
\ee
As shown in \cite{fayyazuddin}, this implies that
$\chi, \phi$ in \eq{zeds} can be written in terms of usual
$2\pi$-periodic angles as
\bea
\chi&& = \eta/M \nn
\phi &&= - \eta/M + \zeta 
\nn
\eta &&\equiv \eta + 2\pi, \zeta \equiv \zeta + 2\pi
\label{periodic}
\eea
In other words  \eq{zequiv} imply the following identifications
for $\phi, \chi$
\begin{eqnarray}
\label{idents}
\chi &&\equiv \chi + {{2\pi}\over M}.{\rm integer} \nn
\phi &&\equiv  \phi + 2\pi.{\rm integer}- {{2\pi}\over M}.{\rm integer}
\end{eqnarray}

\subsection{Semiclassical limit}

The string theory action \eq{action} on AdS$_5 \times (S^5/Z_M)$,
comprising the bosonic terms, is given by
\begin{eqnarray}
\label{action2}
S &&= -{{R^2}\over {4\pi\alpha'}}\int \, d^2\sigma\,
\left[-\cosh^2\rho (\del t)^2 + (\del \rho)^2 +
\sinh^2 \rho (\del \Omega)^2 + \right.
\nn
&& \left. 
\sin^2 \beta (\del \theta)^2 + (\del \beta)^2
+ \cos^2\beta \left( (\del\gamma)^2 + \cos^2\gamma (\del\chi)^2
+ \sin^2\gamma (\del\phi)^2 \right)\right]
\end{eqnarray}
where the identifications \eq{idents} are understood.
$(\del \rho)^2$ means $-(\del_\tau \rho)^2 + (\del_\sigma
\rho)^2$ etc. The range of $\sigma$ is $[0,2\pi]$
and of $\tau$ is $(-\infty,\infty)$.

\def\dele{\del_+}
\def\delf{\del_- }

It is worth writing out the constraints \eq{constraint} explicitly:
\bea
\label{constraint2}
T_{++} =&& -\cosh^2\rho (\dele t)^2 + (\dele \rho)^2 +
\sinh^2 \rho (\dele \Omega)^2 +\sin^2 \beta (\dele \theta)^2 +
\nn
&& (\dele \beta)^2
+ \cos^2\beta \left( (\dele\gamma)^2 + \cos^2\gamma (\dele\chi)^2
+ \sin^2\gamma (\dele\phi)^2 \right)=0
\nn
T_{--} =&& -\cosh^2\rho (\delf t)^2 + (\delf \rho)^2 +
\sinh^2 \rho (\delf \Omega)^2 +\sin^2 \beta (\delf \theta)^2 +
\nn
&& (\delf \beta)^2
+ \cos^2\beta \left( (\delf\gamma)^2 + \cos^2\gamma (\delf\chi)^2
+ \sin^2\gamma (\delf\phi)^2 \right)=0
\eea

Now we seek a solution of
the equations of motion \eq{n-eom},\eq{N-eom} and 
 constraint equations \eq{constraint2} above.

\subsubsection{Collective coordinates}

We note that since $\frac{\del}{\del t},\frac{\del}{\del \chi}$ are
isometries of AdS$_5 \times S^5$ (and of the orbifold), any solution
$\Big(n_a(t(\tau,\sigma),\rho(\tau,\sigma),\ldots)$,
$N^\alpha(\chi(\tau,\sigma), \beta(\tau,\sigma),\ldots)
\Big)$ of the above equations of motion and constraints  will
have the following shift symmetry
\bea
\Big(n_a(t(\tau,\sigma),\rho(\tau,\sigma),\ldots),&& 
N^\alpha(\chi(\tau,\sigma, \beta(\tau,\sigma),\ldots) \Big)
\rightarrow
\nn
&& \Big(n_a(t(\tau,\sigma) + a,\rho(\tau,\sigma),\ldots), 
N^\alpha(\chi(\tau,\sigma)+ b, \beta(\tau,\sigma),\ldots)
\Big)
\label{shift}
\eea
We will further assume that we are considering solutions satisfying
$\rho=0, \beta=0$ (solutions localized near the centre of AdS$_5$,
and close to an equator of $S^5$).
 
As is well-known \cite{Gervais:1974dc}, to take care of the dynamics
along the shift directions we can elevate the parameters $a,b$ to
functions of $\tau$; these are called collective coordinates. Instead
of calling these functions $a(\tau), b(\tau)$ we will permit ourselves
an abuse of notation, calling them $t(\tau), \chi(\tau)$.

The collective coordinate action is easily seen to be
\be
\frac{S_{coll}}{\alpha}= 
\frac{1}{\alpha} \int d\tau\ \left[ (\dot\chi)^2 - 
(\dot t)^2 \right],\; \alpha\equiv \alpha'/R^2
\label{coll-action}
\ee
The constraints \eq{constraint2} become
\be
(\dot\chi)^2 - (\dot t)^2= 4 \dot \chi^+ \dot \chi^- = 0
\ee
where we have introduced
\be
\chi^\pm = \frac12 (\chi \pm t)
\ee
The canonical Hamiltonian
\be
H= \frac1{4} P_+ P_-
\label{can-ham}
\ee
vanishes because of the constraint. Here
\be
P_+ = \frac4{\dot \chi^+},  P_- = \frac4\alpha{\dot \chi^-}
\label{can-mom}
\ee
are the canonical momenta. We have treated $\alpha$ 
as an effective $\hbar$.

The constraint can be made to vanish either by choosing
$P_-=0$ or $P_+=0$. We will choose the  branch $P_-=0$. 
As usual, the constraint generates
redundant motion. This can be fixed by an appropriate gauge
choice. We will choose the gauge 
\be
\chi_-=0.
\label{coll-gauge}
\ee
This, together with the constraint $P_-=0$, become a pair of
second-class constraints.

The quantization of this system is simple. The wavefunctions of the
system which satisfies the constraint are of the form
\be
\psi(t,\chi)= \exp[iP_+ \chi^+/\alpha]
\label{wavefunction}
\ee
Eqns. \eq{idents} imply the following identification for
$\chi^\pm$:
\be
\label{idents2}
\chi^+ \equiv \chi^+ + 2\pi/M,\;
\chi^- \equiv \chi^- + 2\pi/M, 
\ee
which in turn quantizes $P_+$:
\be
P_+ = \alpha.k.M, k= {\rm integer}
\label{mom-quant}
\ee

\gap2
\noindent\underbar{\em Quantized geodesic}
\gap2

The quantum mechanical wavefunction \eq{wavefunction}, together with
the gauge \eq{coll-gauge} is equivalent to the following classical
trajectory in the sigma-model path integral (quantized as above)
\bea
\chi= t = w\tau,
\nn
 \rho=0,  \beta=0
\label{gkp1}
\eea
This represents a spinning particle in the $\chi$ direction
in  $S^5$. The quantization \eq{mom-quant}, through \eq{can-mom}
and \eq{gkp1} implies
\be
w = \alpha.k.M = k. \alpha'M/R^2
\label{w-quant}
\ee

It can be easily checked that the constraints (\ref{constraint2}) are
satisfied, as we anyway ensured in the discussion of the collective
coordinate dynamics.

Thus we see that the classical solution can be consistently discussed
only in the framework of collective coordinates since it involves
$1/\alpha= 1/\hbar$ effects.

In view of Eqn. \eq{w-quant} we note already that the semiclassical
limit $\alpha= \alpha'/R^2 \to 0$ is meaningful only if $M$ is also
sent to infinity in such a way that
\be
\alpha \to 0, M \to \infty, R^2/(\alpha' M) \equiv R_- =
{\rm fixed}
\label{scaling}
\ee

\subsection{Fluctuations}

We now consider the field-theory fluctuations orthogonal to
the collective coordinate motion \eq{gkp1}. 

We will make the gauge choice that 
the $\chi^+$ fluctuation vanishes:
\[
\delta \chi^+=0
\]
In other words the gauge is
\be
\chi^+ = w\tau
\label{full-gauge}
\ee
With this, the constraints
\eq{constraint2}  enable us to solve for $\chi^-$ 
in terms of the other eight variables which we will treat as
independent. One can easily check that the gauge
\eq{full-gauge} can be fixed. The solvability of the 
dependent variable $\chi^-$ in terms of the independent
variables gives an independent justification for the gauge choice
\eq{full-gauge}.

In order to proceed, we will make a semiclassical expansion of the
fluctuations in powers of $\sqrt{\alpha}$. It is easy to see that in
order to have canonically normalized kinetic terms for the independent
variables, we need (since these variables were zero for the
collective motion, we will not explicitly write $\delta \rho$ etc.
and write $\rho$ instead)
\bea
\rho &&= \sqrt\alpha \bar\rho
\nn
\Omega &&= \bar\Omega
\nn
\beta &&= \sqrt\alpha \bar\beta
\nn
\phi && = \bar \phi
\nn
\theta &&= \bar\theta
\nn
\gamma && = \sqrt\alpha \bar \gamma
\label{expansions}
\eea
In view of the classical value \eq{coll-gauge} and the difference
of the two constraints in 
\eq{constraint2}, it is clear that that the dependent variable
$\chi_-$ starts with
\be
\chi_- = \alpha \bar\chi^-,\;
\bar\chi^-= F(\bar\rho,\bar\Omega,\ldots)
\label{chi-detd}
\ee
This equation resolves an important subtlety.  Note that \eq{idents2}
appears to be problematic in the limit \eq{scaling}. However, the fact
that (see \eq{chi-detd}) $\chi^-$ turns out of order 
$ \alpha= \alpha'/R^2$ resolves this problem. In other words,
\eq{idents2} means that the periodicity of
$\bar\chi^-$ is 
\be
\bar\chi^- \equiv \bar\chi^- + 2\pi.m /(M\alpha)
=  2\pi.m  R_-, \; m= {\rm integer}
\label{chi-period}
\ee
so that the period is 
of order one. We have used above the
scaling \eq{scaling}. This is consistent with the second
equation of \eq{chi-detd} where $\bar\chi^-$ is
determined in terms of scaled quantities and hence
of order one. 

\subsubsection{Explicit details of fluctuation calculation}

With the expansions given in \eq{expansions}, the 
constraints \eq{constraint2} become, upto $o(\alpha)$,  
\bea
T_{++} && = 4 \del_+ \chi^+ \del_+ \chi^- + 
\alpha \left[ (\del_+ \vec r)^2 +   (\del_+ \vec y)^2 - w^2 (
\vec r^2 + \vec y^2) \right]=0
\nn
T_{--} && = 4 \del_- \chi^+ \del_- \chi^- + 
\alpha \left[ (\del_- \vec r)^2 +   (\del_- \vec y)^2 - w^2 (
\vec r^2 + \vec y^2) \right]=0
\label{constraint3}
\eea
where $\vec r=(r_1, r_2, r_3, r_4), \vec y= (y_1, y_2,
y_3, y_4)$ are defined as follows 
\bea
&& r_i = \bar \rho \Omega_i,
\nn
&& (y_1, y_2)= \bar \beta (\cos\bar\theta, \sin\bar\theta),
(y_3, y_4)= \bar \gamma (\cos\bar\phi, \sin\bar\phi)
\label{def-r-y}
\eea
We note that the periodicity of $\phi\equiv \bar\phi$ becomes
standard in the $M\to \infty$ limit (see Eqn. 
\eq{idents}). This ensures that $(y_3,y_4)$ represent
a plane just like $(y_1,y_2)$.
 
By using the gauge condition
\eq{full-gauge} we clearly see from the
above constraints that $\chi_-$ must be $O(\alpha)$,
as written earlier in \eq{chi-detd}. 
Using the gauge condition \eq{full-gauge}
and taking the difference of the two constraints we get
\be
\del_\tau \bar \chi^- = - \frac{1}{4 w}\left[
	(\del_\tau \vec r)^2 + (\del_\sigma \vec r)^2 +
(\del_\tau \vec y)^2 + (\del_\sigma \vec y)^2 
-w^2 (\vec r^2 + \vec y^2)\right]
\label{tau}
\ee
\be
\del_\sigma \bar \chi^-= - \frac{1}{4 w}\left[
\del_\tau \vec r.\del_\sigma \vec r + \del_\tau \vec y.\del_\sigma \vec y
\right]
\label{sigma}
\ee
Integrating \eq{sigma}, using $\bar \chi^-(2\pi) -
\chi^-(0) = 2\pi.m. R_-$ according to \eq{chi-period}, and
the value of $w$ from \eq{w-quant}, we
get
\be
4 km + \frac1{2\pi} \int_0^{2\pi} d\sigma \left[
\del_\tau \vec r.\del_\sigma \vec r + \del_\tau \vec y.\del_\sigma \vec y
\right]=0
\ee
This gives the level matching condition of the string
theory.

The Eqn. \eq{tau} can be used to determine both
the charges corresponding to the isometries $\del/\del t,
\del/\del\chi$, namely the energy  $E$ and the angular
momentum $J$ (and consequently $E-J$), as follows. It is easy to see,
e.g. by using the Noether prescription, that $E$ and $J$ are given by
\bea
E &&= \frac1{2\pi \alpha} \int_0^{2\pi}d\sigma 
\cosh^2\rho\ \del_\tau t
\nn
J &&= \frac1{2\pi \alpha} \int_0^{2\pi}d\sigma \cos^2\beta
\cos^2\gamma\ \del_\tau \chi
\eea
By using the expansions \eq{expansions} and \eq{tau},
we get for $E-J$
\be
E-J = \frac1{2w} \int_0^{2\pi}d\sigma \left[
	(\del_\tau \vec r)^2 + (\del_\sigma \vec r)^2 +
(\del_\tau \vec y)^2 + (\del_\sigma \vec y)^2 
+w^2 (\vec r^2 + \vec y^2)\right]
\ee
This gives the spectrum of anomalous dimensions ($\Delta$,
which equals $E$) in the
dual gauge theory \cite{bmn}.

\gap1
\noindent{\em Fermions}
\gap1

We should remark that at this order of fluctuations we should include
fermions. It follows from considerations of supersymmetry that the
eigenvalues of the energy and angular momentum described above remain
the same, however the spectrum becomes supersymmetric.

\gap2
\noindent\underbar{\em Other examples}
\gap2

It is clear from the preceding discussion that our methods are rather
general and can be easily applied to other orbifolds of AdS$_5 \times
S^5$, in particular the ones involving $S^5/(Z_M \times Z_{M'})$ which
preserve ${\cal N}=1$ supersymmetry.  Note that the continuous
isometries of such spaces are the same as those of AdS$_5 \times S^5$.

\section*{Acknowledgments}

We are happy to acknowledge discussions with L. Alvarez-Gaume,
C. Bachas, J. Gauntlett, R. Gopakumar, M.B. Green, C. Hull,
S. Minwalla, C. Nunez and A.K. Raina.  S.R.W. would like to
acknowledge the Isaac Newton Institute for Mathematical Sciences and
Theory Division, CERN for hospitality where part of this work was
done. N.V.S is supported by PPARC Research Assistantship.





\end{document}